\documentclass[12pt,preprint]{aastex}
\usepackage{graphicx,times}             
\usepackage{subfigure}
\usepackage{ulem}
\shorttitle{Discovery of $\gamma$-ray emission of 3C 275.1}

\shortauthors{Liao et al.}

\begin{document}

\title{Discovery of $\gamma$-ray emission from a strongly lobe-dominated quasar 3C~275.1}

\author{Neng-Hui Liao\altaffilmark{1}, Yu-Liang Xin\altaffilmark{1,2}, Shang Li\altaffilmark{1,2},  Wei Jiang\altaffilmark{1,2},  Yun-Feng Liang\altaffilmark{1,2}, Xiang Li\altaffilmark{1,2}, Peng-Fei Zhang\altaffilmark{1,3}, Liang Chen\altaffilmark{4}, Jin-Ming Bai\altaffilmark{5}, Yi-Zhong Fan\altaffilmark{1}}
\affil{$^1$ Key Laboratory of Dark Matter and Space Astronomy, Purple Mountain Observatory, Chinese Academy of Sciences, Nanjing 210008, China}
\affil{$^2$ University of Chinese Academy of Sciences, Yuquan Road 19, Beijing 100049, China}
\affil{$^3$ Department of Physics, Yunnan University, Kunming 650091, China}
\affil{$^4$ Key Laboratory for Research in Galaxies and Cosmology, Shanghai Astronomical Observatory, Chinese Academy of Sciences,80 Nandan Road, Shanghai 200030, China}
\affil{$^5$ Key Laboratory for the Structure and Evolution
of Celestial Objects, Yunnan Observatories, Chinese Academy of Sciences, Kunming 650011, China}
\email{liaonh@pmo.ac.cn (NHL); yzfan@pmo.ac.cn (YZF)}
\begin{abstract}
We systematically analyze the 6-year {\it Fermi}/LAT data of the lobe-dominated quasars (LDQs) in the complete LDQ sample from 3CRR survey and report the discovery of high-energy $\gamma$-ray emission from 3C 275.1. The $\gamma$-ray emission of 3C 207 is confirmed and significant variability of the lightcurve is identified. We do not find statistically significant $\gamma$-ray emission from other LDQs. 3C 275.1 is the known $\gamma$-ray quasar with the lowest core dominance parameter (i.e., $R=0.11$). We also show that both the northern radio hotspot and parsec jet models can reasonably reproduce the $\gamma$-ray data. The parsec jet model, however, is favored by the potential $\gamma$-ray variability at the timescale of months. We suggest that some dimmer $\gamma$-ray LDQs will be detected in the future and LDQs could contribute non-negligibly to the extragalactic $\gamma$-ray background.
\end{abstract}


\section{INTRODUCTION}
Active galactic nuclei (AGNs) powered by accretion of material onto super-massive black holes (SMBHs) are the most luminous and persistent sources of electromagnetic radiation in the Universe. In the orientation-based unified models (Antonucci 1993; Urry \& Padovani 1995), the observed diversity of AGN is ascribed to a few physical parameters, such as the orientation of accretion disk/torus and jet to the observer, the accretion rate, mass and angular momentum of the SMBHs (e.g. Antonucci 1984, Osterbrock \& Pogge 1985; Meier 1999; Ghisellini et al. 2009).

The simple relativistic jet model (Blandford \& Rees 1978) has thus far been reasonably successful in accounting for the primary properties of radio-loud AGNs which constitute about 10 percent of the total. In the unified scheme of radio-loud AGNs, it is postulated that blazars are pole-on-viewed (Blandford \& K\"{o}nigl 1979; Antonucci \& Ulvestad 1985) and hence the radiations of blazars are overwhelming by the luminous and rapidly variable Doppler-boosted jet emission (Ulrich et al. 1997). By comparison, radio-loud AGNs with misdirected jets (i.e., the misaligned AGNs (MAGNs) including radio galaxies and Steep Spectrum Radio Quasars (SSRQs)) exhibit steep radio spectra and bipolar or quasi-symmetrical radio structures. Deboosted radio emissions from mildly relativistic outflows and/or extended radio lobes are significant for MAGNs while the relativistic core radio emissions are dominated for blazars. Ratio of these two components, $R\equiv \rm S_{\rm core}/[S_{\rm total}-S_{\rm core}]$ (which is also called the core dominance parameter, where $S$ represents the observed flux density), is routinely used for classification (Orr \& Browne 1982; Hough \& Readhead 1989). The Lobe dominated quasars (LDQs) are special for SSRQs with $R<1$.

Since the lobe emission should be orientation-unbiased, LDQs are widely used to test the unified scheme of radio-loud AGNs. The anti-correlation between $R$ and the projected linear size of jet $L$ in the complete sample of double-lobed radio quasars is consistent with the expectation that $L$ is foreshortened due to geometrical projection effect at small viewing angles (Hough \& Readhead 1989). Observation of parsec scale morphology of LDQs suggests that no counter jets are seen and the structural variations and the flux variability are mild (Hough et al. 2002). This phenomenon can be well understood via the orientation-dependent relativistic beaming scheme if such objects are viewed at intermediate angles. This argument is supported by the anti-correlation between $R$ and the width of the broad line emissions (e.g.,  Baker \& Hunstead 1995). Thus, the orientation angles of LDQs can be restricted by these approaches, ranging from $\rm 10^{\circ}$ to $\rm 40^{\circ}$ (Aars et al. 2005).

After the successful launch of the {\it Fermi} $\gamma$-ray Space Telescope (Atwood et al. 2009), our understanding of radio-loud AGNs has been revolutionized. In the second {\it Fermi} Large Area Telescope (LAT) source catalog (2FGL, Nolan et al. 2012), the extragalactic $\gamma$-ray sky is dominated by radio-loud AGNs. The vast majority of these sources are blazars (Ackermann et al. 2011). By comparison, MAGNs are not expected to be strong GeV sources (Abdo et al. 2010a). Recently, an analysis of 4-year LAT data shows that the number of significantly detected $\gamma$-ray MAGNs is still handful (Di Mauro et al. 2014). Nevertheless, detecting the $\gamma$-ray emissions from MAGNs is attractive because they offer a different perspective to approach the high-energy phenomena than blazars (Georganopoulos \& Kazanas 2003a). In 2FGL, most $\gamma$-ray detected MAGNs are Fanaroff-Riley type (FR, Fanaroff \& Riley 1974) I radio galaxies whose $\gamma$-ray emissions can be naturally explained by the structured jet radiation models (Georganopoulos \& Kazanas 2003b; Ghisellini et al. 2005). Since they are relatively nearby, $\gamma$-ray emission have been well detected in these sources despite of their large jet inclination angles and small $R$ values. On the other hand, no FR II sources at relatively high redshifts viewed at large angles, such as strongly-lobe-dominated quasars ($R\simeq0.1$), had been reported in 2FGL. The successful detection of   $\gamma$-ray emissions from LDQs is valuable because such data may bring us the structure information of the jets. Not only the  lobes of LDQs are capable of generating strong radio emissions, hotspots in the radio lobes are significant X-ray emitters (e.g. Massaro et al. 2011).  Moreover, the modeling of multi-wavelength spectral energy distributions (SEDs) of these hotspots suggests that they could be potential $\gamma$-ray emitters (e.g. Zhang et al. 2010). The discovery of extended $\gamma$-ray emission from nearby radio galaxy Centaurus A is in support of such a hypothesis (Abdo et al. 2010b). If $\gamma$-ray emissions from the hotspots of some LDQs have been detected, they may consist of non-ignorable part of the extragalactic $\gamma$-ray background (EGB) and may be accelerators of Ultra-High-Energy Comic Rays (Rachen \& Biermann 1993; Massaro \& Ajello 2011).

In this work, we systematically analyze the 6-year {\it Fermi}/LAT data of the complete LDQ sample from the 3CRR survey (Hough \& Readhead 1989) to search for $\gamma$-ray counterparts. This work is organized as follows: In Section 2 the LDQ sample and routines of {\it Fermi}/LAT and Chandra data analysis are introduced. Results of $\gamma$-ray characteristics of the LDQs are reported in Section 3. Finally, in section 4 we summarize our results with some discussions.

\section{THE SAMPLE AND DATA ANALYSES}
\subsection{The Sample}
We analyze the complete LDQ sample (Hough \& Readhead 1989) from the 3CRR survey (Laing et al. 1983). The galactic latitude of each source is derived from the NASA Extragalactic Database (NED)\footnote{http://ned.ipac.caltech.edu/}. All sources except 3C 175 and 3C 181 are well above the Galactic plane (i.e., $\rm |b| > 20^{\circ}$). Therefore, the $\gamma$-ray analysis of these LDQs does not suffer from significant background contamination.

All sources considered in this work exhibit steep radio spectra and FR II radio morphology. The redshifts range from $0.31$ to $2.02$ with a median value of $1.11$. These LDQs are well detected in radio, optical and X-ray bands. Radio structures at both parsec scale from VLBI observations and kilo-parsec scale from VLA observations are available (Hough et al. 2002; Gilbert et al. 2004; Fernini 2014). All LDQs have also been covered by the optical spectrophotometric observations (e.g. Aars et al. 2005). Therefore, the masses of the central SMBHs are constrained by the width of the broad line emissions (McLure et al. 2006). The archival X-ray data of the sources except 3C 175 and 3C 336 are available (Belsole et al. 2006; Hardcastle et al. 2006; Wilkes et al. 2013). For 3C 275.1 without a known detailed X-ray spectrum we have performed an individual analysis of its archival Chandra data. The multi-wavelength properties of these LDQs are summarized in Table 1. Not only for the central core, the hotspot emissions of a few LDQs in the sample are also resolved by the infrared, optical and X-ray observations (Cheung et al. 2005; Massaro et al. 2011; Werner et al. 2012).

\subsection{LAT Data Analysis}

The {\it Fermi}/LAT (Atwood et al. 2009) is a pair-conversion $\gamma$-ray telescope sensitive to photon energies greater than 20 MeV. It has a large peak effective area ($\sim$8000 $\rm cm^{2}$ for 1 GeV photons) monitoring $\simeq$ 2.4 sr of the full sky with angular resolution (68\% containment radius) better than $\rm 1^{\circ}$ at 1 GeV. In its routine survey mode, LAT performs a complete and uniform coverage of the sky in every 3 hours.

The latest {\it Pass 7} Reprocessed data used in our analysis are collected in the energy range of $0.1-100$ GeV  during the first 6-year operation (i.e., from 2008 August 4th to 2014 August 4th). Photon events belonging to the {\it SOURCE} class have been taken into account. The updated standard {\it ScienceTools} software package version {\it v9r33p0} together with the instrument response functions of ${\rm P7REP\_V15}$ are adopted throughout the data analysis. The galactic diffuse model we take is ${\rm gll\_iem\_v05\_rev1.fit}$ and  the isotropic diffuse emission template is taken as ${\rm iso\_source\_v05\_rev1.txt}$  \footnote{http://fermi.gsfc.nasa.gov/ssc/data/access/lat/BackgroundModels.html}. The entire data set is filtered with {\it gtselect} and {\it gtmktime} tasks following the standard analysis threads\footnote{http://fermi.gsfc.nasa.gov/ssc/data/analysis/scitools/}.

The {\it binned} likelihood algorithm implemented in the {\it gtlike} task has been adopted to extract the flux and spectrum. The region-of-interest (ROI) of each LDQ is taken to be a $20^{\circ} \times 20^{\circ}$ box centered at its radio position. Such a size is to match the broad point spread function (PSF) of 100~MeV photons. The data are binned into 30 logarithmically distributed energy bins and $100 \times 100$ spatial bins with size of $0.2^{\circ}$.  All sources from 2FGL (Nolan et al. 2012) within $20^{\circ}$ of the radio position are included. The flux and spectral parameters of sources within the ROI together with normalization factors of the two diffuse backgrounds are set free, while parameters of other sources are fixed to be that reported in the 2FGL. Firstly, we add a presumed $\gamma$-ray source corresponding to the LDQ into the initial background model generated from {\it make2FGLxml.py}\footnote{http://fermi.gsfc.nasa.gov/ssc/data/analysis/user}. Position of the $\gamma$-ray source is assumed to be the same as the radio position of the LDQ and its spectral template is utilized as Power-law with an index $\Gamma_{\rm ph}$  fixed as 2.5  (Throughout this work we refer to a spectral index $\alpha$ as the energy index such that $F_{\nu}\propto\nu^{-\alpha}$, corresponding to a photon index $\Gamma_{ph}=\alpha+1$), consistent with the nominal spectral index of MAGNs (Kataoka et al. 2011). The Test Statistic (TS, Mattox et al. 1996) value of the central $\gamma$-ray source is calculated following the standard approach. To better check the robustness of the signal, we make a $16^{\circ} \times 16^{\circ}$ scale residual TS map with each pixel of $0.2^{\circ}$. If any $\gamma$-ray excess with TS value over 25 appears in the TS map, we add new sources with Power-law spectral template into the background model to account for these excesses, and then fit the updated background model. If no statistically-significant signal from the central source has been founded, the $2\sigma$ flux upper limit is presented.

\subsection{Chandra Data Analysis}
3C 275.1 was observed by Chandra on June 2, 2001 with an exposure time of 24.76 ks. The observation was performed by ACIS-S567 detector in Faint mode without grating and the TE exposure mode. \texttt{CIAO} 4.6 with the \texttt{CALDB} version 4.5.9 is utilized. And \texttt{XSPEC} version 12.8.2 is adopted to fit the spectrum. The data set is obtained from the Chandra Data Archive\footnote{http://cda.harvard.edu/chaser (ObsID: 2096)}. The standard Chandra analysis threads are followed. Target events are extracted from a circular region with a radius of 4.5 arcsec while the background events are from a nearby circle of the same radius, giving a net count rate of $\rm 0.189\pm0.003$. The non-negligible pile-up affection ($\rm \simeq15\%$) has been corrected. In spectral analysis, absorption column density is fixed as $\rm N_{H}=1.99\times10^{20}$~$\rm cm^{-2}$, consistent with previous studies (e.g. Crawford \& Fabian 2003). The spectrum has been grouped to require at least 30 counts $\rm bin^{-1}$ so that the result of $\chi^{2}$ statistical analysis is ensured to be valid. A single Power-law model provides a well description of the data, $\chi^{2}/d.o.f$(97/103), giving the unabsorbed 0.5-8.0 keV flux of $\rm 1.29^{+0.02}_{-0.07}\times10^{-12}$ erg $\rm cm^{-2}$ $\rm s^{-1}$ and $\Gamma_{ph}$ of $\rm 1.53\pm0.08$. The X-ray data are also evenly divided into five sub-energy bins, for the SED modeling in following section.

\section{RESULTS}
Results of the 6-year $\gamma$-ray data analysis are presented in Table 2. Significant $\gamma$-ray excesses with TS values $\gtrsim$ 50 (the corresponding significance is $\simeq 7\sigma$) around the radio positions of LDQs only display in three sources, including 3C 14, 3C 207 and 3C 275.1. We note that the $\gamma$-ray emission of 3C 207 has been reported before (Abdo et al. 2010a; Nolan et al. 2012). Tentative excesses with TS values $\simeq 10$ are found for 3C 208 and 3C 212 and more data are needed to confirm the detection. For majority of the LDQs in our sample, the TS values of the assumed central $\gamma$-ray sources are less than one.  In addition to the fits of the 6-year $\gamma$-ray data together, for LDQs without hint of significant $\gamma$-ray emission, we also perform a detailed variability analysis to check whether there could be significant emission in a shorter period. For such a purpose we divide the whole 6-year $\gamma$-ray data into ten time bins. If the TS value in one time bin is above 5, we then divide it into monthly bins. In such an approach we do not find any excess with ${\rm TS\geq 25}$  but a tentative excess with a ${\rm TS \simeq 18}$ for 3C 191 is found within 20 days. Detailed analyses of 3C 275.1, 3C 14 and 3C 207 are introduced below.

\subsection{Detecting $\gamma$-ray emission of 3C 275.1}
3C 275.1 is a well-studied bright extragalactic radio source (Bridle \& Perley 1984). It is identified as a quasar by optical spectroscope observation (Hintzen 1984). The source is the first quasar found at the centre of rich cluster of galaxies (Hintzen \& Stocke 1986). A parsec-scale radio image of 15~GHz exhibits a typical core-jet structure with jet extending toward northwest (Hough et al. 2002). A 5~GHz VLA image shows that one-sided jet links to the north edge-brightened radio lobe with a S-shaped bend, while the opposite lobe does not clearly connect to the core component (Stocke et al. 1985; Gilbert et al. 2004). The hotspot in the north radio lobe is not only detected in radio observations, but also resolved by {\it Spitzer}, {\it Hubble Space Telescope (HST)} and {\it Chandra} (Crawford \& Fabian 2003; Cheung et al. 2005; Werner et al. 2012).

A point-like $\gamma$-ray excess with ${\rm TS\simeq 70}$ appears in the center of the residual map. However, such a $\gamma$-ray excess might be artificial suffered by uncertainty of the Galactic background diffuse emission or nearby bright neighbors.\footnote{See the definition of data  flags in 2FGL (Nolan et al. 2012), especially for {\it flags} 1-5.} The excess locates at a high Galactic latitude ($\rm \textit{l}\simeq 79^{\circ}$), so the affect from the uncertainty of Galactic diffuse emission is negligible. On the other hand, in both 1FGL and 2FGL there is no strong $\gamma$-ray source within $\rm 3^{\circ}$ from the excess (Abdo et al. 2010c; Nolan et al. 2012). In addition to the $\gamma$-ray sources reported before, we find other four background excesses in the residual map and the closest $\gamma$-ray source is $\rm 2.8^{\circ}$ away from the central excess. Its TS value ($\simeq 25$) is considerably lower than the central one. Considering that the angular resolution of LAT increases rapidly with the energy in the sub-GeV energy range, we make a $10^{\circ} \times 10^{\circ}$ TS map for the 0.6-100 GeV $\gamma$-rays (Note that for the 0.6 GeV photons the 95\% PSF containment angle is $\leq 3^{\circ}$) with the background model in which the target is not included, as shown Figure 1. The central $\gamma$-ray excess is still statistically significant (i.e., ${\rm TS>25}$), suggesting that the detection of $\gamma$-ray emission is robust.

Localization of the central excess is performed by the {\it gtfindsrc} task and we have a $\gamma$-ray position of R.A. $\rm 191.041^{\circ}$ and DEC. $\rm 16.3485^{\circ}$, with a 95\% confidence level (C. L.) error radius of $\rm 0.118^{\circ}$ ($\rm 424^{\prime\prime}$). As a high Galactic latitude source, a radio-loud AGN is supposed to be its ideal counterpart. We seek the potential counterpart through the SIMBAD database\footnote{http://simbad.u-strasbg.fr/simbad/}. 3C 275.1 is found to be the only radio-loud AGN within the 95\% C.L. error radius (the angular separation from the best fit position is $\rm 212^{\prime\prime}$). The second nearest AGN is NGC 4651, which is just out of the 95\% error radius but is a normal spiral galaxy harboring a low-ionization emission line nuclear. The lack of strong starburst activity indicates that NGC 4651 is probably not capable of generating significant $\gamma$-ray emission (Ackermann et al. 2012). The nearest blazar candidate from the $\gamma$-ray position is BZB J1244+1616 (Massaro et al. 2009) but the angular separation is so large ($\rm 567^{\prime\prime}$) that the association is highly disfavored. Motivated by these facts, we conclude that the central significant $\gamma$-ray excess is from 3C 275.1. \footnote{When preparing for this manuscript, we were informed that the {\it Fermi}/LAT collaboration reported the preliminary list of MAGN of 3LAC in the 5th {\it Fermi} Symposium, in which 3C 275.1 was included.}

The best-fit Power-law spectrum for 3C 275.1 is
\begin{equation}
 \frac{dN}{dE}=(8.72\pm1.23)\times10^{-13}(\frac{E}{\rm 809.24~MeV})^{-(2.52\pm0.12)},
\end{equation}
and an integrated flux is $\rm (11.20\pm2.53)\times10^{-9}$ photons $\rm cm^{-2}$ $\rm s^{-2}$. At a redshift $z=0.557$, the isotropy $\gamma$-ray luminosity in the energy range of $0.1-100$ GeV is $\rm (8.17\pm1.19)\times10^{45}$ erg $\rm s^{-1}$. Note that in this work we take a $\Lambda$CDM cosmology with $H_{0}=70~{\rm km~ s^{-1}~Mpc^{-1}}$, $\Omega_{\rm m}=0.3$, and $\Omega_{\Lambda}=0.7$ (Komatsu et al. 2011).  The $\gamma$-ray SED is extracted by dividing the whole data into 7 sub-energy bins. A power-law fit gives an acceptable description to the SED, which well agrees with the entire fit (see Figure 2). A $\gamma$-ray light curve consisting of 10 time bins has been also extracted, as shown in Figure 3. As TS values of major time bins are around 10, we fix the spectral indexes to the value of the average fit. The large statistic errors make further variability analysis impossible. However, different from other time bins, the TS value of the last time bin is relatively high ($\simeq 20$). And considering its 1$\sigma$ statistic errors, the last bin is ``well" above the average flux level, which indicates possible $\gamma$-ray variability at timescale of months.

\subsection{3C 14: no reliable $\gamma$-ray emission}
3C 14 is a high redshift ({\it z} = 1.469) steep spectral radio quasar with an extremely lobe-dominated ($R$=0.01) radio morphology (Laing et al. 1983). The 8.4 GHz VLA radio image shows an asymmetric structure with the jet-linked southeastern lobe while no counter jet for the north side lobe (Fernini 2014). No significant structure variation ($\beta_{app} \sim 0$) has been found in parsec scale from VLBA observations, and the inclination angle is suggested to be $39^{\circ}$ (Aars et al. 2005).

Similar to 3C 275.1, in the $\gamma$-ray analysis we find a point-like significant $\gamma$-ray excess (${\rm TS\simeq 47}$) around the radio position of 3C 14. Since 3C 14 is also a high Galactic latitude ($\rm \textit{l}\simeq 49^{\circ}$) source and the nearest $\gamma$-ray source is $\rm 2.3^{\circ}$ away, we re-fit the 6-year LAT data in the energy range from 800 MeV to 100 GeV and the central excess is still significant (TS $\geq 25$), suggesting that the excess is intrinsic. Furthermore, the $\gamma$-ray localization gives a position of R.A. $\rm 9.1793^{\circ}$ and DEC. $\rm 18.6215^{\circ}$ with a 95\% C.L. error radius $\sim 0.179^{\circ}$. In the SIMBAD database 3C 14 is the only known radio-loud AGN within the 95\% C.L error radius. However, there is a nearby radio source CRATES J003659+183202 (J0036+1832) characterized by the flat radio spectrum (Healey et al. 2007).

To derive the properties of this $\gamma$-ray excess, we perform a fit to 6-year LAT data by utilizing a power-law spectral template, which gives an averaged photon flux of $(8.55\pm2.20)\times10^{-9}$ ph $\rm cm^{-2}$ $\rm s^{-1}$ with a spectral index of $\Gamma_{\rm ph}=2.41\pm0.13$. Interestingly, in the 7-month time bin analysis we have ${\rm TS<4}$  in most bins except in one that has ${\rm TS\simeq 140}$, as shown in Figure 4. More detailed variability analysis in such an outburst phase suggests that the emission was mainly from one month (Dec. 2012). Individual fit for the $\gamma$-ray photons in this short epoch gives the $\gamma$-ray flux of $(1.53\pm0.25)\times10^{-7}$ ph $\rm cm^{-2}$ $\rm s^{-1}$ and $\Gamma_{\rm ph}=2.16\pm0.11$, with a TS value of 157. Since the $\gamma$-ray emission in such an intense radiation phase suffered from little background contamination, the location error can be significantly reduced. Indeed, the new $\gamma$-ray position is found to be R.A. $\rm 9.2736^{\circ}$ and DEC. $\rm 18.5441^{\circ}$, with a 95\% C.L. error radius of $\rm 0.108^{\circ}$. Surprisingly, 3C 14 falls out from location radius (see Figure 5). On the other hand, the flat spectral radio source, CRATES J0036+1832 is still within the location radius. It is thus reasonably to speculate that the counterpart of the $\gamma$-ray source is CRATES J0036+1832 and 3C 14 is probably not able to radiate significant $\gamma$-ray emission. A 2-day time bin light curve is extracted and the peak flux is $(3.43\pm1.06)\times10^{-7}$ ph $\rm cm^{-2}$ $\rm s^{-1}$, which is roughly 40 times higher than the 6-year average flux. Such a violent $\gamma$
-ray variability behavior is typically observed for blazars (e.g. Liao \& Bai 2015), rather than MAGNs, supporting our $\gamma$-ray localization result.

\subsection{$\gamma$-ray emission from 3C 207}
3C 207 at a redshift $z=0.684$ has been reported as a $\gamma$-ray emitter in the first and second LAT AGN catalogs (Abdo et al. 2010d; Ackermann et al. 2011). It shows a marginally lobe-dominant ($R=0.45$) radio morphology (Hough \& Readhead 1989). Its radio core flux density is the second strongest in the sample (Hough et al. 2002). Besides of the significant variability of the radio core, structure variation with $\beta_{\rm app}\simeq10$ is reported, which is the typical behavior of blazars (Hough et al. 2002; Lister et al. 2013) and an inclination angle $\approx 11^{\circ}$ is inferred (Aars et al. 2005). Not only high energy emission from the nucleus has been observed, X-ray emission from hotspot of the northern radio lobe has also been detected (Brunetti et al. 2002).

Detailed $\gamma$-ray analysis for 3C 207 had been performed in Abdo et al. (2010a). There are three AGNs falling into the 95\% C.L. error radius while 3C 207 has the highest association possibility ($P=99\%$). In the first two years, its $0.1-100$ GeV photon flux was $(2.35\pm0.37)\times10^{-8}$ ph $\rm cm^{-2}$ $\rm s^{-1}$, the photon spectrum index was $\Gamma_{\rm ph}=2.36\pm0.11$ and the TS value was about 64. On the other hand, the fit to the entire 6-year LAT data gives the photon flux of $(1.36\pm0.24)\times10^{-8}$ ph $\rm cm^{-2}$ $\rm s^{-1}$ and $\Gamma_{\rm ph}=2.60\pm0.11$. The resulting TS value is about 70, which does not change considerably. We also derive the $\gamma$-ray location by analyzing the 6-year data. We confirm that the radio position of 3C 207 is still within the 95\% error radius. However it is at the edge of error radius (see Figure 6). Nevertheless, 3C 207 remains as the only known RL-AGN within the $\gamma$-ray localization error radius. The 7-month time bin light curve indicates that 3C 207 has significant variability at timescale of months. The flux of the first 7 months is $5\sigma$ separated from the 6-year average flux (see Figure 7). The significant $\gamma$-ray variability and the high apparent speed of the ejected knots are self-consistent with its relatively small inclination angle.

\section{Summary and Discussions}
In the analysis of the first 15 months Fermi/LAT data, M87 ($z$ = 0.004) has the lowest {\it R} value (0.05) among the FRI $\gamma$-ray emitters while 3C 207 has the lowest {\it R} value (0.45) for FR IIs with significant $\gamma$-ray emission  (Abdo et al. 2010a).  Interestingly 3C 207 was the unique LDQ in 2LAC (Ackermann et al. 2011). More recently the $\gamma$-ray emission from BLRG Pictor A, a strongly lobe-dominant source ($R$ = 0.08) with FR II morphology, has been detected by {\it Fermi}-LAT  (Kataoka et al. 2011; Brown \& Adams 2012). However, it is the known nearest $\gamma$-ray FRII source at the redshift $z$ = 0.035. Such a low redshift accords with the typical redshift of FRIs rather than FRIIs. The corresponding $\gamma$-ray luminosity of Pictor A is just $\rm \simeq 10^{43}$ erg $\rm s^{-1}$, similar with those of FRI radio galaxies. Our systematic analysis of the 6-year {\it Fermi}-LAT data for the selected LDQs is the first try for this extreme class of MAGNs.   A GeV source likely associating with 3C 207 has been confirmed and strong variability is identified.  Furthermore, significant $\gamma$-ray emission from 3C 275.1 is discovered and its core dominance ($R$ = 0.11) is lower than that of 3C 207. This is the first time to detect $\gamma$-ray emission from a luminous strongly-lobe-dominant FR II quasar. Now the lowest $R$ values for $\gamma$-ray detected radio galaxies (M87) and Quasars (3C 275.1) are comparable. Luminous $\gamma$-ray emission ($\rm \simeq 10^{46}$ erg $\rm s^{-1}$) at large viewing angles ($\rm \sim20^{\circ}$) might indicate new $\gamma$-ray origin, such as electron-positron pair jets (Ghisellini 2012a). Therefore, it is intriguing to study the $\gamma$-ray radiation mechanism of 3C 275.1.

Due to the limited angular resolution of LAT, the $\gamma$-ray emitting site can not be directly inferred. In principle it could be from the northern hotspot where an S-shape structure has been observed that may be indicative of intense interaction between jet and nearby galaxy (Stocke et al. 1985; Gilbert et al. 2004).  And $\gamma$-ray emission from this hotspot has been predicted in the modeling of radio to X-ray emission (e.g. Zhang et al. 2010). On the other hand, multiwavelength observations suggest that the jet component contributes to the core flux from radio to X-rays. Significant variability of core flux with a factor $\simeq2$ at timescale of years has been observed in radio and optical bands (Sandage et al. 1965; Hough et al. 2002; Schneider et al. 2010). And the spectrum of the core X-ray emission is very hard ($\Gamma \simeq 1.5$). Such a hard spectrum can not be explained by the classic disk-corona scenario and hence X-ray jet component should be taken into account (Dou \& Yuan 2008). Moreover, the optical core emission has a linear polarization degree of $(4.2\pm1.5)\%$ (Wills et al. 2011), indicating that polarized synchrotron jet emission component is not negligible.

We adopt the classic homogeneous lepton radiation model to calculate both the hotspot and core jet emissions (see Figure 9), in which both the synchrotron and inverse Compton scattering (IC) processes have been taken into account. The input parameters are summarized in Table 3. The radiating electrons are assumed to follow a broken Power-law distribution and both self-synchrotron absorption and Klein-Nishina effects have been addressed  (detailed description please see Liao et al 2014). In the hotspot scenario, synchrotron plus synchrotron self-Compton (SSC) without significant relativistic beaming ($\delta=1$) and synchrotron plus IC/CMB process under mild relativistic condition are considered. Interestingly, in the SSC model the observed $\gamma$-ray spectrum well accords with the natural extension of the X-ray data, see Figure 9a. And physical parameters from SED modeling are consistent with those found in Zhang et al. (2010). In contrast, the IC/CMB model can not well describe the X-ray and $\gamma$-ray emissions simultaneously (see also Zhang et al. 2010). In the core radiation scenario, synchrotron plus SSC+IC/IR model under a mild relativistic condition is adopted. Such an approach is motivated by the VLBA Polarimetry observations of several LDQs that polarization of their parsec scale jets are higher than their cores, indicating that the parsec scale jets point closer to our line of sight and hence are less obscured by a Faraday screen (Aars \& Hough 2005). The external photon field is assumed to consist of 1 eV IR photons that could be from the hot dust. The energy density is fixed as $\rm 3\times10^{-4}$ erg $\rm cm^{-3}$ in the rest frame, in agreement with that given in Ghisellini et al. (2012b). Both the hotspot and core scenarios can well reproduce the $\gamma$-ray emission of 3C 275.1. However, in view of its potential $\gamma$-ray variability at timescale of months, the core scenario may be favored.

The location of the $\gamma$-ray emission is important for understanding the energy dissipation of MAGNs. Very high energy $\gamma$-ray flare of M87 has been found to be simultaneous to an X-ray flare associated with the innermost knot of the jet (Harris et al. 2006), or a radio flare produced within 10 Schwarzschild radii from the central SMBH (Acciari et al. 2009). In addition, significant $\gamma$-ray emissions both from the extensive structure and the central core of Centaurus A have been detected (Abdo et al. 2010b; Abdo et al. 2010e).  Nevertheless, there is increasing evidence that the location of $\gamma$-ray emission of MAGN is near the central engine. The variability of $\gamma$-ray emissions from MAGNs suggests that the contribution of $\gamma$-ray emission from the radio hotspot/lobe is unimportant (e.g. Grandi et al., 2012; Torresi \& Grandi 2013). For 3C 275.1, there is tentative evidence for $\gamma$-ray variability at timescale of months.  The $\gamma$-ray emission and the the multi-wavelength hotspot radiation of Pictor A are hard to explain self-consistently (Brown \& Adams 2012). Moreover, $\gamma$-ray radiation locations of 3C 111 and 3C 120  are only $\sim$0.5 pc from the central SMBH, as constrained by simultaneous multi-wavelength observation (Grandi et al., 2012; Tanaka et al., 2015). 3C 275.1 is also an ideal target for simultaneous multi-wavelength variability study since the $\gamma$-ray radiation location of SSRQs is still unclear. The phenomenon that the ejection of super-luminous knot is coupled with a dip in X-ray light curve of some nearby BLRGs (e.g. Chatterjee et al. 2011) may also be detected for SSRQs.

Doppler beaming effect is believed to be crucial for producing $\gamma$-ray emission of AGN.  Many pieces of evidence, such as rapid $\gamma$-ray variations at timescale of hours and minutes (e.g. Liao \& Bai 2015), suggest that $\gamma$-ray emissions of blazars are from the pc scale inner jets pointing to us. However, MAGNs tend to have milder $\gamma$-ray variability and slower knot apparent speeds than blazars, consistent with their observed large jet inclination angles. And FR I radio galaxies are less $\gamma$-ray luminous than their parent population of BL Lac objects (Abdo et al. 2010a), see Figure 9, in accordance with the unified scheme of radio-loud AGN.  Corresponding to the observed structured jets of FRI radio galaxies and BL Lac objects, Meyer et al. (2011) proposed a gradient Lorentz scenario that observers at different viewing angles see different speed portions of the jet and hence the $\gamma$-ray spectra of FRI radio galaxies are softer than BL Lac objects. On the other hand,  for the FRII sources and in particular the SSRQs, they occupy the typical FSRQs region (Abdo et al. 2010a). The isotropic $\gamma$-ray luminosity of 3C 275.1 is $\simeq$ $10^{46}$ erg $\rm s^{-1}$, again in the FSRQs region (see also Figure 9). SSRQs are suggested to be more strongly Doppler boosted when comparing with radio galaxies.  Indeed, significant $\gamma$-ray variabilities of 3C 380 and 3C 207 are detected, and their observed super-luminal motion are over $10c$, in accordance with their relatively small jet inclination angle of $\sim$$\rm 10^{\circ}$ (Torresi \& Grandi 2013). It is somewhat puzzling that 3C 275.1, a source with a large jet inclination angle ($\sim$$\rm 20^{\circ}$) according to its relatively low $R$ value, is $\gamma$-ray luminous ($\simeq$ $10^{46}$ erg $\rm s^{-1}$). However, if its pc jet is bending to us, the actual angle between the $\gamma$-ray emission region of 3C 275.1 and the line of sight can be smaller, possibly $\sim$$\rm 10^{\circ}$.  Except the SSRQs,  there are a few FR II radio galaxies detected by {\it Fermi}/LAT, e.g. 3C 111 and Pictor A. They are less $\gamma$-ray luminous than SSRQs, as shown in Figure 9, probably due to the milder relativistic beaming effect by their relatively larger jet inclination angle of $\sim$$\rm 20^{\circ}$ (Jorstad et al., 2005). Significant difference of the $\gamma$-ray spectral index between the FRII radio galaxies and FRII SSRQs are not found, in accordance with the argument in Meyer et al. (2011) that the gradient Lorentz effect does not hold for FRII and FSRQs.

A tight correlation between radio and $\gamma$-ray luminosities has been found for blazars  (e.g. Padovani et al. 1993; Ghirlanda et al. 2011). And a connection between core radio flux at 5~GHz and the $\gamma$-ray luminosity is reported for MAGNs (Di Mauro et al. 2014). As shown in Figure 10, 3C 275.1 follows such a relationship, with Spearman rank-order correlation coefficient of 0.95 and $p$-value of $\rm 3\times10^{-7}$. This result is consistent with the preference that its $\gamma$-ray is contributed by the core. The upper limits for other LDQs have been also plotted in the plane, and majority of them are above the correlation line, indicating that the number of $\gamma$-ray LDQ could increase when the exposure accumulates. Therefore, like other MAGNs, LDQs could contribute non-ignorable part of the EGB.

\acknowledgements
We appreciate the helpful suggestions from the anonymous referee that led to a substantial improvement of this work. This research has made use of data obtained from the High Energy Astrophysics Science Archive Research Center (HEASARC), provided by $\rm NASA^{\prime}$s Goddard Space Flight Center. This research has also made use of the NASA/IPAC Extragalactic Database which is operated by the Jet Propulsion Laboratory, California Institute of Technology, under contract with the National Aeronautics and Space Administration. This research makes use of the SIMBAD database, operated at CDS, Strasbourg, France. This work was supported in part by 973 Programme of China under grant 2013CB837000, National Natural Science of China under grants 11361140349, 11433009, 11133006 and 11233006, and the Foundation for Distinguished Young Scholars of Jiangsu Province, China (No. BK2012047). NHL thanks Xin-Wu Cao for his advice. NHL and PFZ thanks Shan-shan Weng and Teng Liu for the suggestions of the Chandra data analysis.

\begin{figure}
\centering
\includegraphics[scale=0.6]{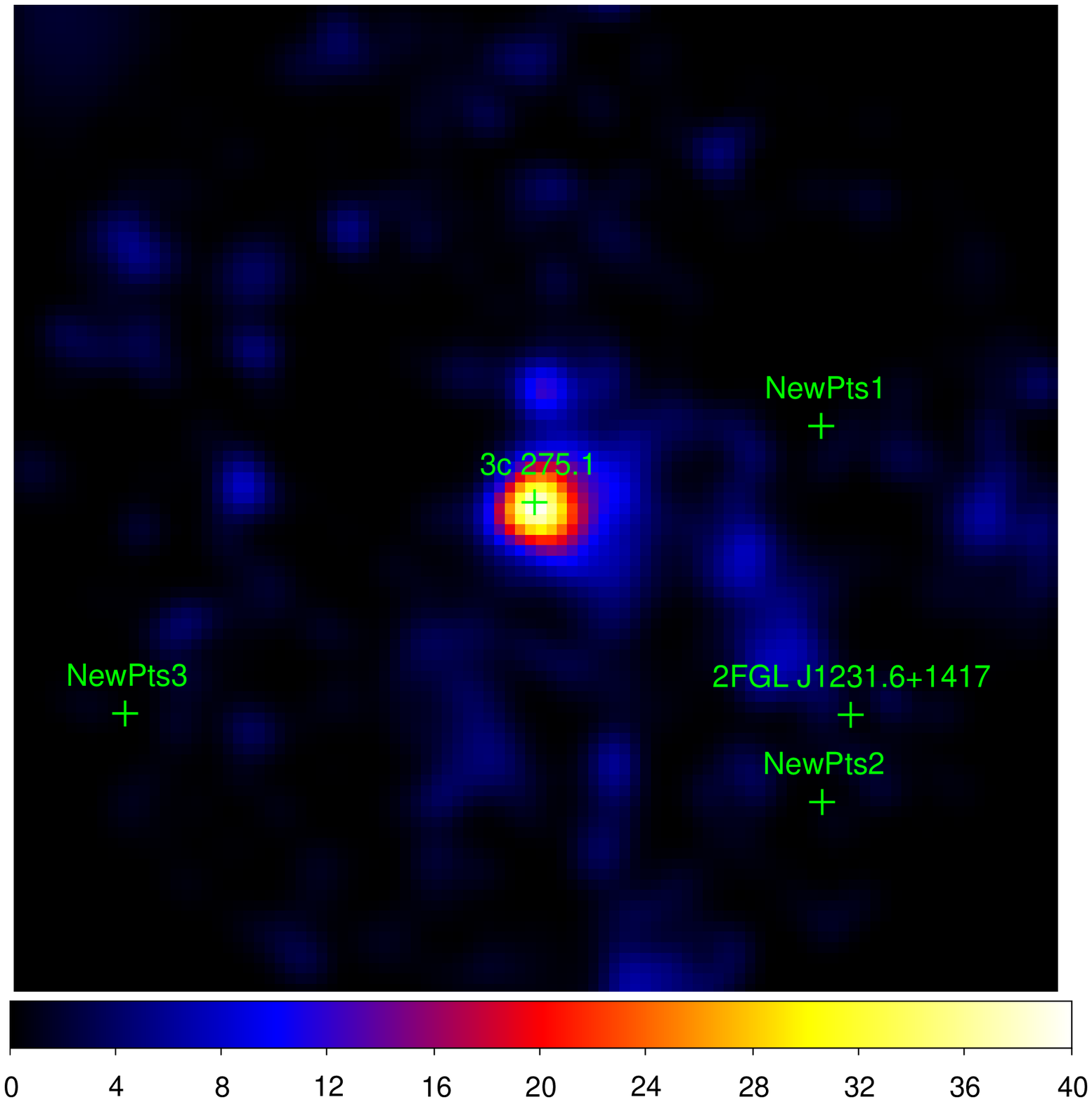}
\caption{TS map of photons from 600 MeV to 100 GeV for $10^{\circ} \times 10^{\circ}$ region centered at 3C 275.1. The diffuse backgrounds, 2FGL and additional sources are subtracted. TS value of the central excess corresponding to 3C 275.1 is consistent with {\it gtlike} analysis. Beside of the target, $\gamma$-ray neighbors within $5^{\circ}$ are listed. The map is smoothed with $\sigma$=$0.3^{\circ}$ Gaussian function.}
\label{Fig.1}
\end{figure}

\begin{figure}
\centering
\includegraphics[scale=0.4]{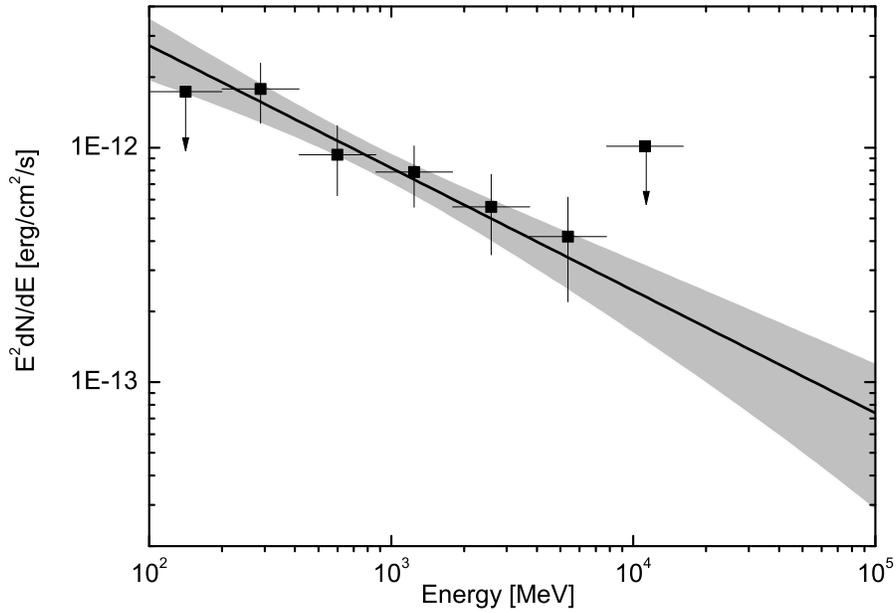}
\caption{$\gamma$-ray SED of 3C 275.1. The solid line represents the best fit of the entire 6-year data. The shadow  is the 1$\sigma$ uncertainty area. The black squares are the individual fits for sub-energy bins.}
\label{Fig.2}
\end{figure}

\begin{figure}
\centering
\includegraphics[scale=0.35]{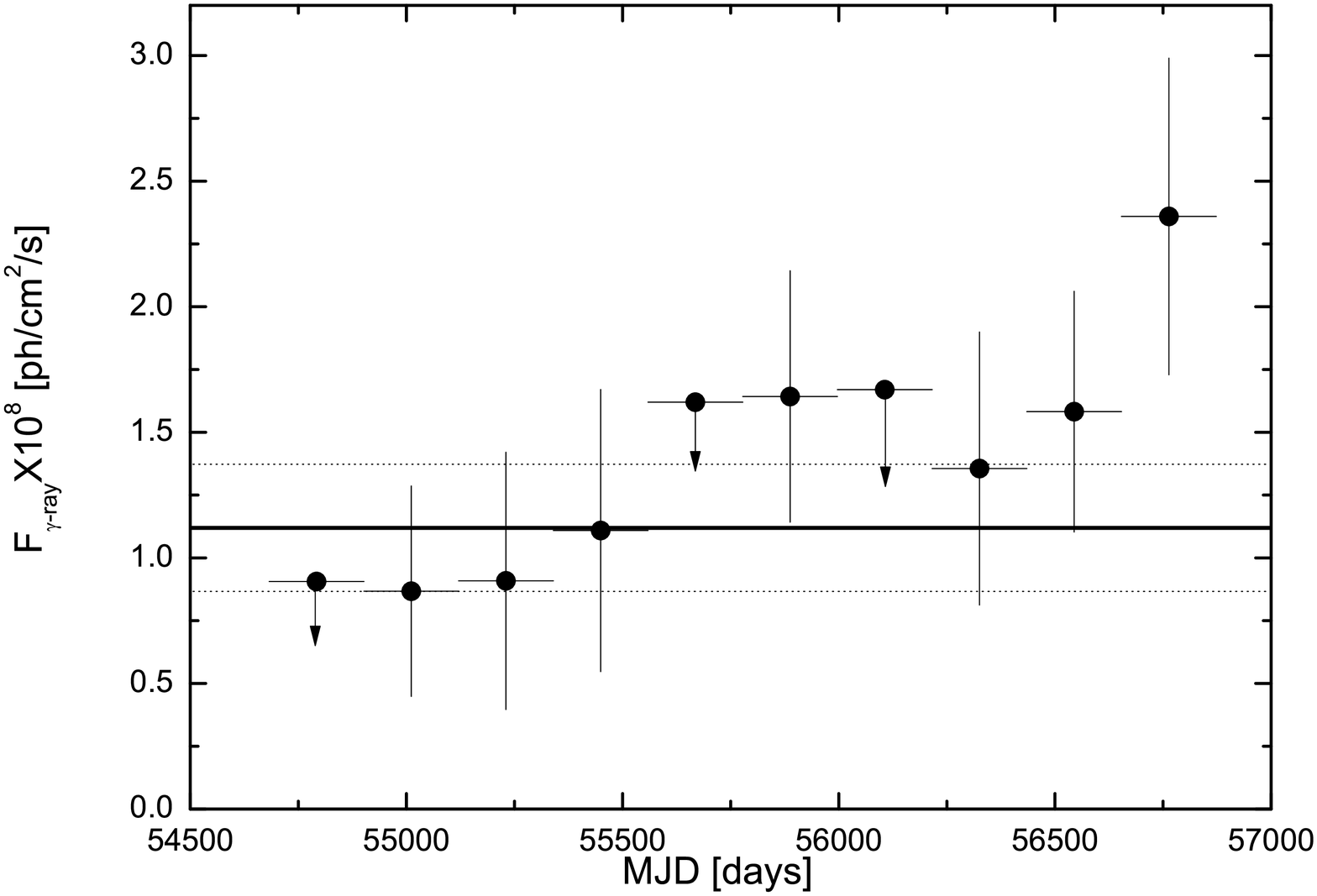}
\caption{$\gamma$-ray light curve of 3C 275.1. Upper limits are derived for time bins with a TS value smaller than 4. The solid line is the 6-year average flux whose 1$\sigma$ flux error is marked by the two dotted lines.}
\label{Fig.3}
\end{figure}

\begin{figure}
\centering
\includegraphics[scale=0.35]{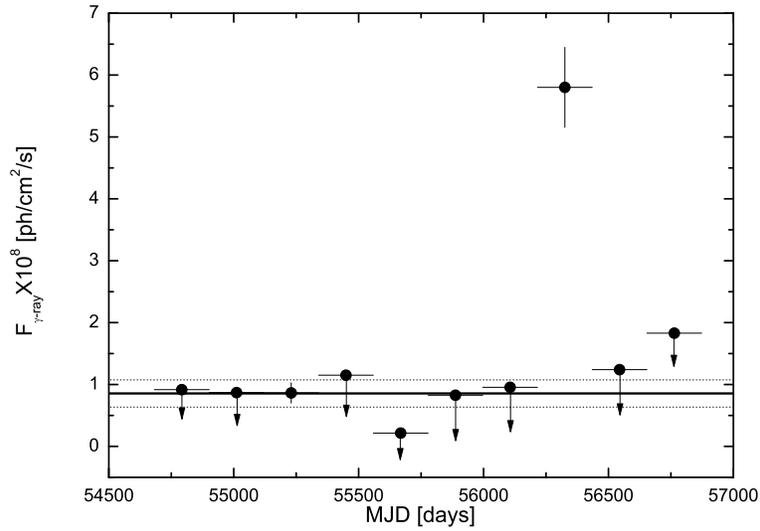}
\caption{Light curve of $\gamma$-ray excess around the radio position of 3C 14. Upper limits are derived for time bins with a TS value smaller than 4. The solid line is the 6-year average flux whose 1$\sigma$ flux error is marked by the two dotted lines.}
\label{Fig.4}
\end{figure}

\begin{figure}
\centering
\includegraphics[scale=0.35]{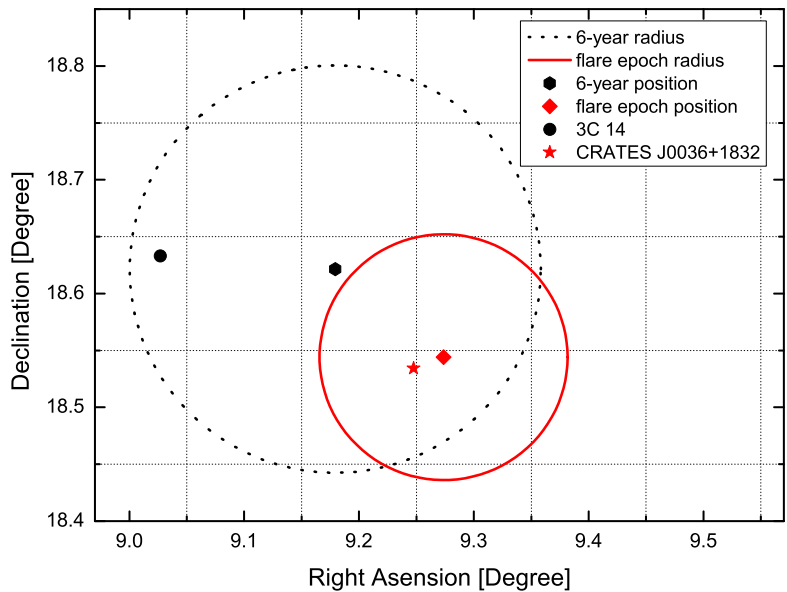}
\caption{$\gamma$-ray localization of the $\gamma$-ray excess around the radio position of 3C 14. Evidently the radio position of 3C 14 falls out from the error radius of the flare epoch though it is within the error radius found in the 6-year data analysis.}
\label{Fig.5}
\end{figure}

\begin{figure}
\centering
\includegraphics[scale=0.35]{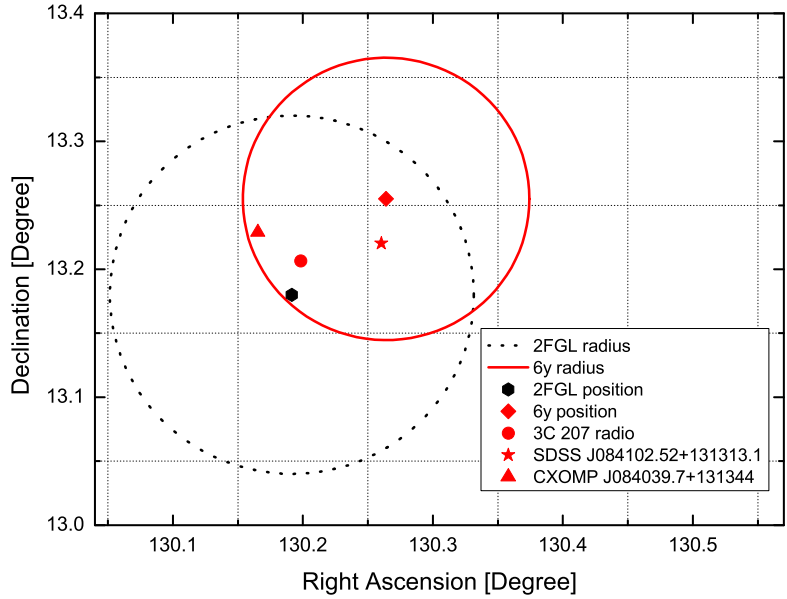}
\caption{$\gamma$-ray localization of 3C 207 by analyzing 6-year LAT data, together with its $\gamma$-ray location and 95\% error radius listed in 2FGL (Nolan et al. 2010). The radio position of 3C 207 is within both the first 2-year and 6-year $\gamma$-ray location radii. However, it is not the nearest counterpart in the 6-year LAT data localization analysis.}
\label{Fig.6}
\end{figure}

\begin{figure}
\centering
\includegraphics[scale=0.35]{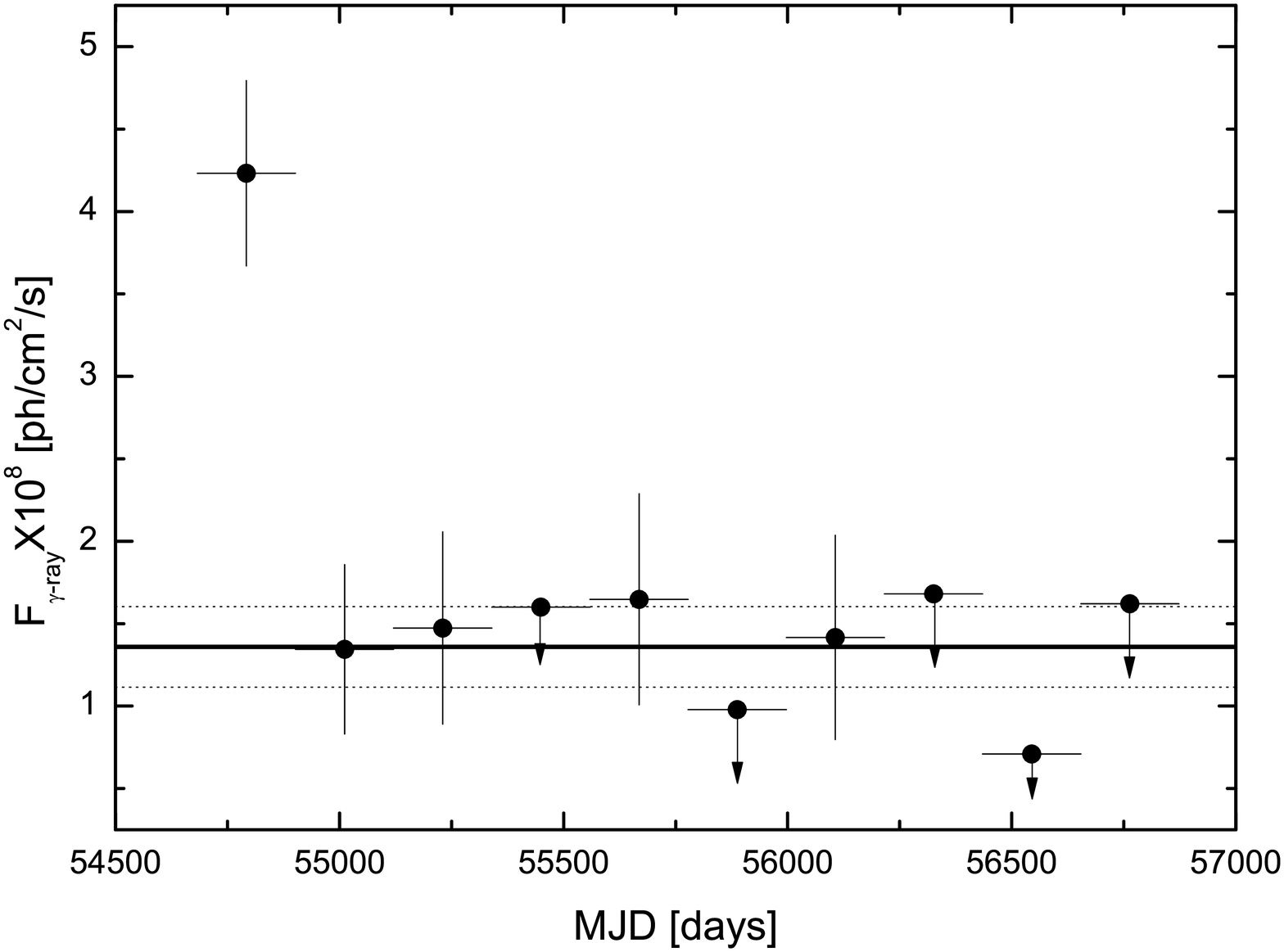}
\caption{$\gamma$-ray light curve of 3C 207. Upper limits are derived for time bins which TS values are lower than 4. The solid line is the 6-year average flux whose 1$\sigma$ flux error is marked by the two dotted lines.}
\label{Fig.7}
\end{figure}

\begin{figure}
\centering
\subfigure[]{
\includegraphics[scale=0.246]{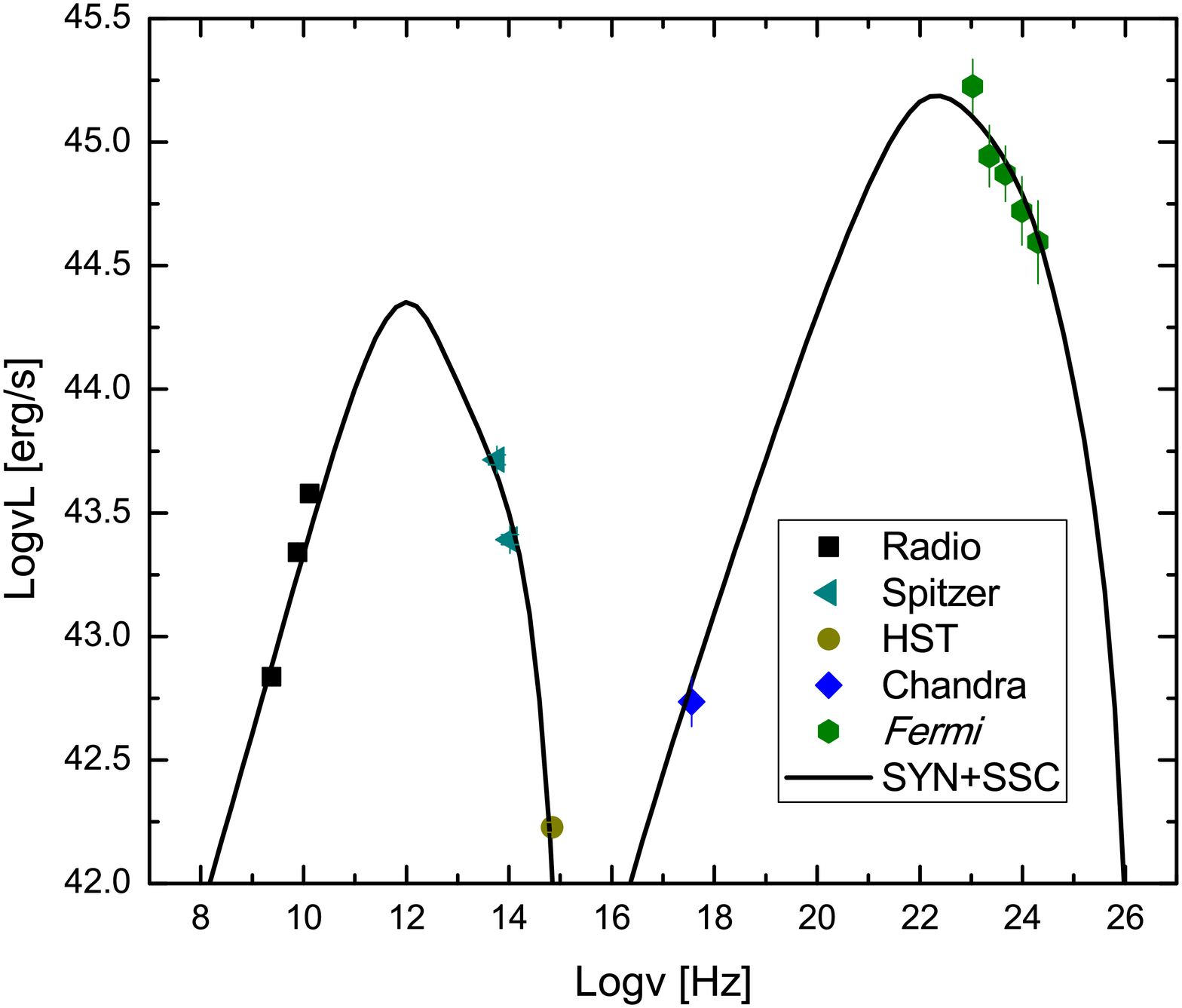}}
\subfigure[]{
\includegraphics[scale=0.25]{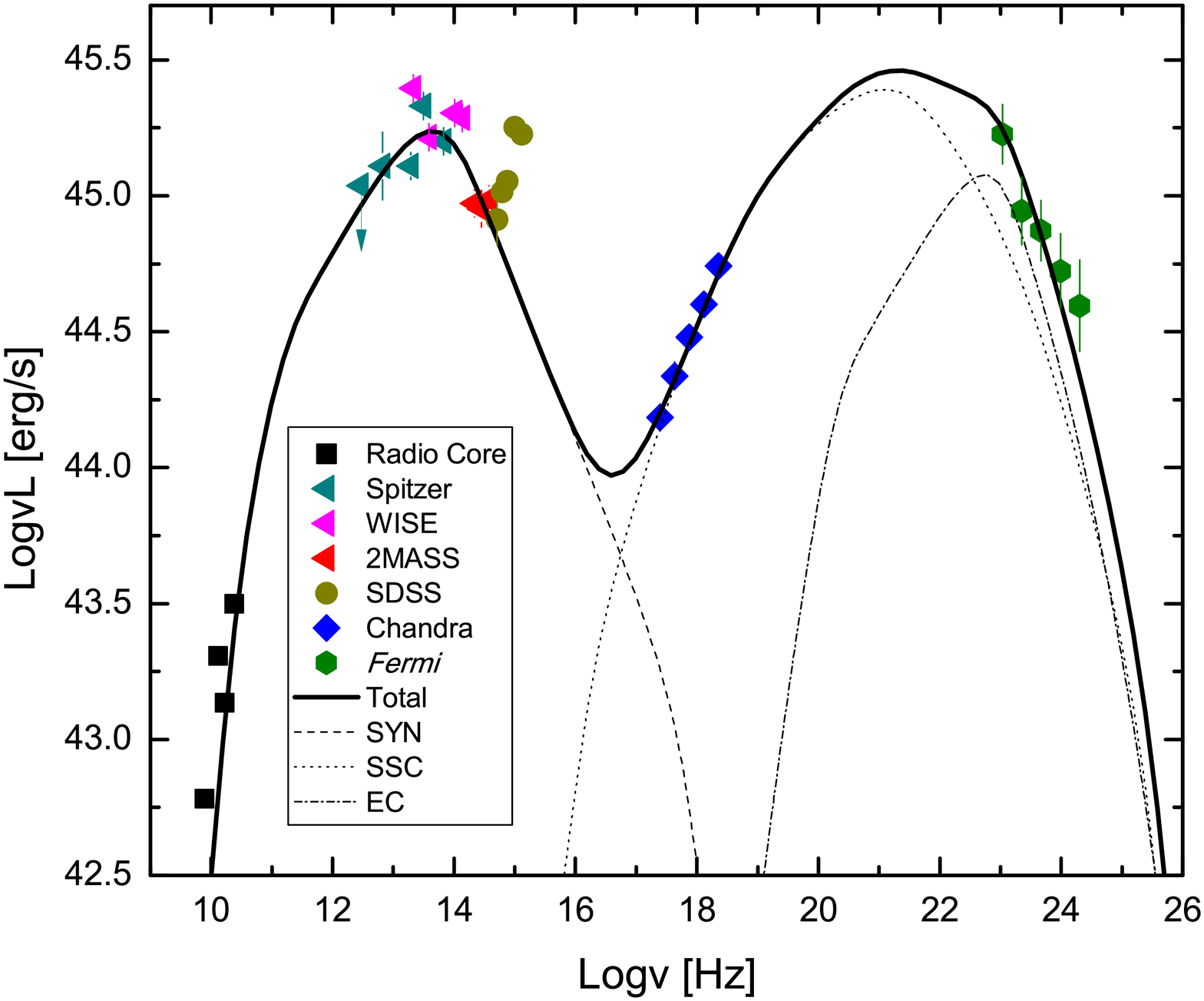}}
\caption{(a)SED of the northern hotspot emission of 3C 275.1, together with the synchrotron plus SSC modeling; (b)SED of the core emission of 3C 275.1, together with the synchrotron plus SSC+EC modeling. Collected multi-wavelength data: radio data (Laurent-Muehleisen et al. 1997; Hough et al. 2002); IR data (Spitzer, Cleary et al. 2007, Werner et al. 2012; WISE, Wright et al. 2010; 2MASS, Skrutskie et al. 2006); optical data (Cheung et al. 2005, Schneider et al. 2010); Chandra X-ray data (Crawford \& Fabian 2003). The corrections for the interstellar extinction and the color excess of optical data have been adopted from Schlegel et al. (1998). Jumps in the core SED are not considered for modeling, because they are probably thermal emissions from the dust torus and accretion disk.}
\label{figure8}
\end{figure}

\begin{figure}
\centering
\includegraphics[scale=0.35]{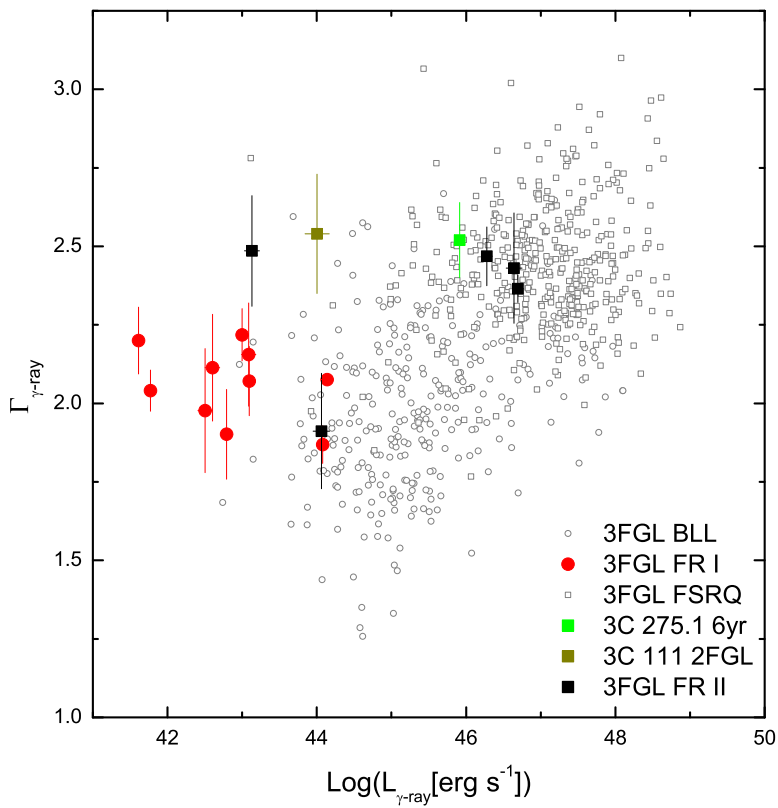}
\caption{Plane of the 0.1-100 GeV $\gamma$-ray luminosity and $\gamma$-ray spectral index for MAGNs, together with 3LAC blazars. Data of blazars and MAGNs listed in 3LAC (Ackermann et al. 2015) are obtained from their data website (http://www.asdc.asi.it/fermi3lac/). Note that $\gamma$-ray emission of 3C 111 in the 3LAC may be contaminated by a recently merging strong $\gamma$-ray neighbor so its 2LAC values are adopted in this plane. Redshift modifications between the emitted and observed energies have been performed for all sources. }
\label{Fig.9}
\end{figure}

\begin{figure}
\centering
\includegraphics[scale=0.35]{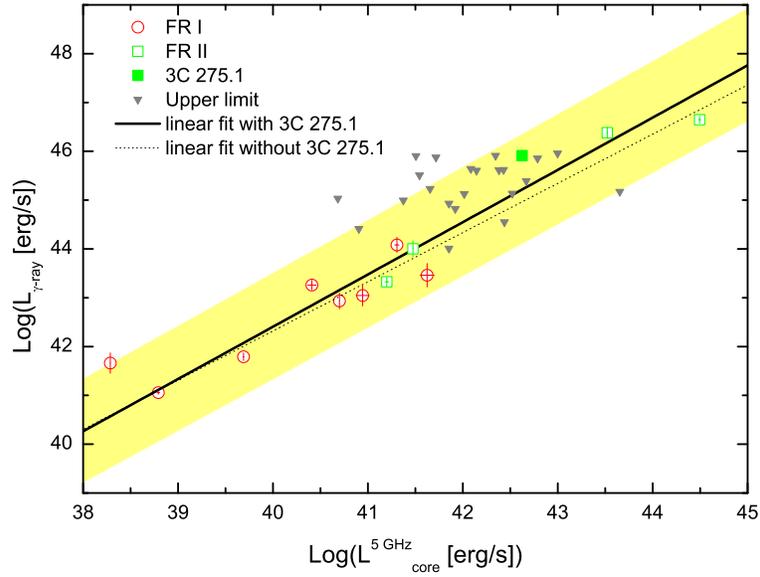}
\caption{Plane of observed $\gamma$-ray luminosity and radio core luminosity at 5 GHz for the MAGNs. The hollow symbols represent the known $\gamma$-ray MAGNs which data are taken from  Di Mauro et al. (2014). The filled green square is the location of 3C 275.1 and the grey triangles are upper limits for most of the LDQs. The solid line corresponds to best linear fit for $\gamma$-ray MAGNs including 3C 275.1, together with 90\% uncertainty area (yellow area). The dashed line is the best fit for $\gamma$-ray MAGNs without 3C 275.1 presented in Di Mauro et al. (2014).}
\label{Fig.10}
\end{figure}

\begin{deluxetable}{lcccccccccccc}
\scriptsize
\tablenum{1} \tablewidth{0pt}
\tablecaption{Multiwavelength Properties of LDQs in the Complete LDQ Sample from 3CRR Survey}
\tablehead{ \colhead{3CR name} &\colhead{\it b\tablenotemark{a}} &\colhead{\it z\tablenotemark{b}} &\colhead{\it R} &\colhead{$\rm S_{178 MHz}$} &\colhead{$\rm S_{Core 5 GHz}$} &\colhead{$m_{v}$} &\colhead{$\rm f_{1-keV}$\tablenotemark{c}} &\colhead{$\rm M_{BH}$\tablenotemark{d}}
}

\startdata
3C 9 &$-46.533^{\circ}$ &2.018 &0.004 &546 &5 &18.21 &7.19(0.38) &9.8 \\[3pt]
3C 14 &$-44.088^{\circ}$ &1.469 &0.010 &606 &11 &20.0 &21.12(2.48) &9.4 \\[3pt]
3C 47 &$-40.698^{\circ}$ &0.425 &0.05 &1092 &72 &18.1 &162.25(43.28) &9.2 \\[3pt]
3C 68.1 &$-23.988^{\circ}$ &1.238 &0.0007 &824 &1 &19.5 &9.20(4.21) &9.9\\[3pt]
3C 175 &$10.080^{\circ}$ &0.768 &0.024 &655 &25 &16.6 &$\rm \cdot\cdot\cdot$ &9.9 \\[3pt]
3C 181 &$14.630^{\circ}$ &1.387 &0.005 &655 &6 &18.92 &13.79(1.50) &9.6 \\[3pt]
3C 190 &$21.841^{\circ}$ &1.195 &0.081 &814 &100 &20.32 &8.24(0.49) &8.7 \\[3pt]
3C 191 &$20.901^{\circ}$ &1.956 &0.026 &600 &35 &18.4 &7.61(0.42) &9.7 \\[3pt]
3C 204 &$35.512^{\circ}$ &1.112 &0.044 &338 &28 &18.21 &27.06(2.28) &9.5 \\[3pt]
3C 205 &$36.896^{\circ}$ &1.536 &0.018 &665 &24 &17.62 &21.55(1.23) &9.6 \\[3pt]
3C 207 &$30.139^{\circ}$ &0.684 &0.49 &1240 &510 &18.15 &19.13(2.66) &8.5\\[3pt]
3C 208 &$33.158^{\circ}$ &1.11 &0.048 &536 &51 &17.42 &20.28(1.99) &9.4 \\[3pt]
3C 212 &$34.504^{\circ}$ &1.048 &0.12 &884 &150 &19.06 &46.59(1.68) &9.2 \\[3pt]
3C 215 &$37.249^{\circ}$ &0.411 &0.039 &407 &20 &18.27 &55.86(4.11) &8.3 \\[3pt]
3C 245 &$56.300^{\circ}$ &1.029 &0.99 &1600 &910 &17.29 &35.02(1.01) &9.4 \\[3pt]
3C 249.1 &$38.550^{\circ}$ &0.311 &0.12 &775 &100 &15.72 &189.81(85.60) &9.3 \\[3pt]
3C 263 &$49.744^{\circ}$ &0.652 &0.10 &1009 &169 &16.32 &45.26(7.02) &9.1 \\[3pt]
3C 268.4 &$71.404^{\circ}$ &1.396 &0.046 &596 &50 &18.42 &22.86(2.44) &9.8 \\[3pt]
3C 270.1 &$80.639^{\circ}$ &1.516 &0.11 &943 &190 &18.6 &14.61(0.83) &9.0 \\[3pt]
3C 275.1 &$79.115^{\circ}$ &0.557 &0.11 &890 &130 &19.0 &31.2(3.2) &8.3\\[3pt]
3C 334 &$41.108^{\circ}$ &0.555 &0.23 &720 &180 &16.41 &43.77(5.32) &9.7 \\[3pt]
3C 336 &$42.102^{\circ}$ &0.927 &0.024 &760 &29 &17.47 &$\rm \cdot\cdot\cdot$ &9.2 \\[3pt]
3C 351 &$36.382^{\circ}$ &0.371 &0.006 &1202 &8 &15.28 &42.15(3.65) &9.5 \\[3pt]
4C 16.49 &$24.006^{\circ}$ &1.296 &0.010 &400 &9 &18.4 &13.41(1.48) &9.8 \\[3pt]
3C 432 &$-22.825^{\circ}$ &1.805 &0.009 &361 &8 &17.96 &8.06(0.59) &10.1 \\[3pt]
\enddata
\tablenotetext{a}{Galactic latitudes are adopted from NED database.}
\tablenotetext{b}{The redshifts are taken from Hough et al. (2002). {\it R} values, radio flux densities (in unit of mJy) are from Hough \& Readhead (1989). Apparent V band magnitudes are taken from Aars et al. (2005).}
\tablenotetext{c}{X-ray data are collected from the literature (Belsole et al. 2006; Hardcastle et al. 2006; Wilkes et al. 2013), in scale of $\rm 10^{-14}$ erg $\rm cm^{-2}$ $\rm s^{-1}$, except 3C 175 and 3C 336.}
\tablenotetext{d}{Black hole mass data in logarithmic form are adopted from McLure et al. (2006).}

\end{deluxetable}
\clearpage

\begin{deluxetable}{lccccccccc}
\scriptsize
\tablenum{2} \tablewidth{0pt}
\tablecaption{Results of Analyzing the Entire 6-year LAT Data}
\tablehead{ \colhead{3CR name} &\colhead{TS value} &\colhead{flux\tablenotemark{a}} &\colhead{flux error} &\colhead{$\Gamma_{ph}$\tablenotemark{b}} &\colhead{$\Gamma_{ph}$ error}
}

\startdata
3C 275.1 &69.7 &11.20 &2.53 &2.52 &0.12 \\[3pt]
3C 207 &70.2 &13.59 &2.44 &2.60 &0.11 \\[3pt]
\noalign{\smallskip} \hline \noalign{\smallskip}
3C 14\tablenotemark{c} &47.2 &8.59 &2.13 &2.41 &0.13 \\[3pt]
\noalign{\smallskip} \hline \noalign{\smallskip}
3C 9 &$\rm <$1 &1.18 &$\rm \cdot\cdot\cdot$ &2.5f &$\rm \cdot\cdot\cdot$ \\[3pt]
3C 47 &$\rm <$1 &3.62 &$\rm \cdot\cdot\cdot$ &2.5f &$\rm \cdot\cdot\cdot$ \\[3pt]
3C 68.1 &$\rm <$1 &0.84 &$\rm \cdot\cdot\cdot$ &2.5f &$\rm \cdot\cdot\cdot$ \\[3pt]
3C 175 &$\rm <$1 &1.53 &$\rm \cdot\cdot\cdot$ &2.5f &$\rm \cdot\cdot\cdot$  \\[3pt]
3C 181 &$\rm <$1 &5.20 &$\rm \cdot\cdot\cdot$ &2.5f &$\rm \cdot\cdot\cdot$ \\[3pt]
3C 190 &$\rm <$1 &2.05 &$\rm \cdot\cdot\cdot$ &2.5f &$\rm \cdot\cdot\cdot$ \\[3pt]
3C 191 &$\rm <$1 &1.58 &$\rm \cdot\cdot\cdot$ &2.5f &$\rm \cdot\cdot\cdot$ \\[3pt]
3C 204 &3.3 &4.11 &$\rm \cdot\cdot\cdot$ &2.5f &$\rm \cdot\cdot\cdot$ \\[3pt]
3C 205 &$\rm <$1 &2.23 &$\rm \cdot\cdot\cdot$ &2.5f &$\rm \cdot\cdot\cdot$ \\[3pt]
3C 208 &12.6 &7.66 &$\rm \cdot\cdot\cdot$ &2.5f &$\rm \cdot\cdot\cdot$ \\[3pt]
3C 212 &11.9 &7.40 &$\rm \cdot\cdot\cdot$ &2.5f &$\rm \cdot\cdot\cdot$ \\[3pt]
3C 215 &6.8 &5.74 &$\rm \cdot\cdot\cdot$ &2.5f &$\rm \cdot\cdot\cdot$ \\[3pt]
3C 245 &$\rm <$1 &1.59 &$\rm \cdot\cdot\cdot$ &2.5f &$\rm \cdot\cdot\cdot$ \\[3pt]
3C 249.1 &$\rm <$1 &1.02 &$\rm \cdot\cdot\cdot$ &2.5f &$\rm \cdot\cdot\cdot$ \\[3pt]
3C 263 &$\rm <$1 &8.59 &$\rm \cdot\cdot\cdot$ &2.5f &$\rm \cdot\cdot\cdot$ \\[3pt]
3C 268.4 &$\rm <$1 &2.68 &$\rm \cdot\cdot\cdot$ &2.5f &$\rm \cdot\cdot\cdot$  \\[3pt]
3C 270.1 &5.8 &5.12 &$\rm \cdot\cdot\cdot$ &2.5f &$\rm \cdot\cdot\cdot$ \\[3pt]
3C 334 &2.1 &4.46 &$\rm \cdot\cdot\cdot$ &2.5f &$\rm \cdot\cdot\cdot$ \\[3pt]
3C 336 &$\rm <$1 &1.71 &$\rm \cdot\cdot\cdot$ &2.5f &$\rm \cdot\cdot\cdot$ \\[3pt]
3C 351 &$\rm <$1 &1.84 &$\rm \cdot\cdot\cdot$ &2.5f &$\rm \cdot\cdot\cdot$ \\[3pt]
4C 16.49 &$\rm <$1 &1.24 &$\rm \cdot\cdot\cdot$ &2.5f &$\rm \cdot\cdot\cdot$ \\[3pt]
3C 432 &$\rm <$1 &3.23 &$\rm \cdot\cdot\cdot$ &2.5f &$\rm \cdot\cdot\cdot$ \\[3pt]
\enddata
\tablenotetext{a}{$\gamma$-ray fluxes and their 1$\sigma$ errors, together with the 2$\sigma$ upper limits, are in scale of $\rm 10^{-9}$ ph $\rm cm^{-2}$ $\rm s^{-1}$.}
\tablenotetext{b}{2.5f means that the spectral index is fixed as 2.5.}
\tablenotetext{c}{Further $\gamma$-ray localization and variability analysis suggest that the counterpart of this $\gamma$-ray source is a flat spectral radio source CRATES J0036+1832 rather than 3C 14, more details are given in section 3.2.}

\end{deluxetable}

\begin{deluxetable}{lccccccccc}
\scriptsize
\tablenum{3} \tablewidth{0pt}
\tablecaption{Input Parameters of the SED Models\tablenotemark{a}}
\tablehead{ \colhead{Model} &\colhead{$\rm p_{1}$} &\colhead{$\rm p_{2}$} &\colhead{$\gamma_{br}$} &\colhead{K} &\colhead{B} &\colhead{$\delta$} &\colhead{R\tablenotemark{b}}
}

\startdata
hotspot (SSC) &1.6 &3.9 &$\rm 9.3\times10^{4}$ &$\rm 1.8\times10^{-4}$ &$\rm 2.7\times10^{-5}$ &1 &$\rm 5.8\times10^{21}$ \\[3pt]
Core (IC/IR) &2.2 &4.2 &$\rm 4.8\times10^{3}$ &$\rm 3.5\times10^{3}$ &0.3 &2.4 &$\rm 6.0\times10^{17}$ \\[3pt]
\enddata

\tablenotetext{a}{$\rm p_{1,2}$ are the indexes of the broken-powerlaw radiative electron distribution; $\gamma_{br}$ is the break energy of the electron distribution; K is the normalization of the particle number density; B is the magnetic field strength; $\delta$ is the Doppler boosting factor and R is the radius of the emission blob.}

\tablenotetext{b}{In the hotspot modeling, the radius of the emission hotspot is constrained by the {\it HST} observation, $\simeq0.2$ arsec (Cheung et al. 2005). In the core modeling, the radius of the emission region is limited by R$\leq ct_{var}\delta(1+z)^{-1}$, where $\rm t_{var}$ is set as 150 days, to be consistent with its potential $\gamma$-ray variability.}

\end{deluxetable}

\end{document}